\documentclass[aip,jcp,reprint,floatfix]{revtex4-1}  
\pdfoutput=1

\usepackage[T1]{fontenc}
\usepackage[utf8x]{inputenc}

\usepackage[pdftex, bookmarks, pdffitwindow=false, pdfstartview=FitH, pdfdisplaydoctitle, colorlinks, plainpages=false, pdftitle={Local order parameters for use in driving homogeneous ice nucleation with all-atom models of water}, pdfpagelabels, hypertexnames, citecolor={blue}, pdflang={en}, hyperfootnotes=false, breaklinks]{hyperref}

\usepackage{graphicx}
\usepackage{amsmath,amssymb}

\usepackage{fixmath}

\usepackage[slantedGreek]{mathptmx}
\usepackage[Symbol]{upgreek}
\usepackage[basic,italic]{mathastext}
\usepackage{helvet}

\usepackage{setspace}

\usepackage[stretch=15,shrink=15,step=3]{microtype}
\usepackage{cmap}

\def\avg#1{\ensuremath{\left\langle #1 \right\rangle}}
\def\abs#1{\ensuremath{\left| #1 \right|}}

\makeatletter
\renewcommand\section{%
  \@startsection
    {section}%
    {1}%
    {\z@}%
    {0.4cm \@plus 0.5ex \@minus .1ex}%
    {0.2cm}%
    {%
     \normalfont
     \small
     \sffamily
     \bfseries
     \raggedright
    }%
}
\renewcommand\subsection{%
  \@startsection
    {subsection}%
    {2}%
    {\z@}%
    {.4cm \@plus 0.5ex \@minus .1ex}%
    {.2cm}%
    {%
     \normalfont
     \small
     \sffamily
     \bfseries
     \raggedright
    }%
}
\makeatother

\begin{document}

\title{Local order parameters for use in driving homogeneous ice nucleation with all-atom models of water}

\author{Aleks Reinhardt}
\author{Jonathan P.~K.~Doye}
\email[Author~for~correspondence.~Electronic~mail:~]{jonathan.doye@chem.ox.ac.uk}
\affiliation{Physical and Theoretical Chemistry Laboratory, Department of Chemistry, University of Oxford, Oxford, OX1 3QZ, United Kingdom}
\author{Eva G.~Noya}
\affiliation{Instituto de Qu\'{i}mica F\'{i}sica Rocasolano, CSIC, Calle Serrano 119, E-28006 Madrid, Spain}
\author{Carlos Vega}
\affiliation{Departamento de Qu\'{i}mica F\'{i}sica, Facultad de Ciencias Qu\'{i}micas, Universidad Complutense de Madrid, E-28040 Madrid, Spain}
\date{22 October 2012}

\begin{abstract}
We present a local order parameter based on the standard Steinhardt--Ten Wolde approach that is capable both of tracking and of driving homogeneous ice nucleation in simulations of all-atom models of water. We demonstrate that it is capable of forcing the growth of ice nuclei in supercooled liquid water simulated using the TIP4P/2005 model using overbiassed umbrella sampling Monte Carlo simulations. However, even with such an order parameter, the dynamics of ice growth in deeply supercooled liquid water in all-atom models of water are shown to be very slow, and so the computation of free energy landscapes and nucleation rates remains extremely challenging.
\end{abstract}



\maketitle

\section{Introduction}
It is well-known that substances cooled below their thermodynamic freezing point do not necessarily freeze, especially when they are very pure. Homogeneous nucleation is a kinetically disfavoured process; according to classical nucleation theory, a critical cluster must spontaneously form in the supercooled liquid before crystallisation can proceed.\cite{Laaksonen1995, Auer2005, Anwar2011b} The kinetic barrier, in the framework of classical nucleation theory, arises from a competition between a favourable bulk free energy difference between the phases and the unfavourable formation of an interface between the phases. The presence of a free energy barrier to nucleation makes homogeneous nucleation a rare event.

Homogeneous nucleation has been studied using computer simulations in a range of systems;\cite{Anwar2011b, Sear2012} however, the nucleation of ice, despite being of fundamental interest,\cite{Hagen1981, *Toner1990, *Oxtoby1992, *Pruppacher1995, *Baker1997, *Koop2000, *Zachariassen2000, *Debenedetti2003, *Zachariassen2004, *Benz2005, *Hegg2009, *Spichtinger2010, *JohnMorris2011} is still proving to be difficult to simulate successfully despite the apparent simplicity of the process.  The crystallisation of ice has been studied in a large number of simulations;\cite{Svishchev1994, Svishchev1996, Yamada2002, Matsumoto2002, Nada2003, Radhakrishnan2003b, Radhakrishnan2003, Nada2005, Carignano2005, Fernandez2006, Vrbka2006, Vrbka2007, Carignano2007, Bauerecker2008, Quigley2008, Brukhno2008, Pluharova2010,  Moore2010, Kastelowitz2010, Moore2010b, Pereyra2011, Weiss2011, Rozmanov2011, Moore2011b, Li2011, Pirzadeh2011b, Pirzadeh2011, GonzalezSolveyra2011, Nada2011, Yan2011, Reinhardt2012, Johnston2012, Malkin2012, Cox2012, Yan2012, Rozmanov2012b} nevertheless, whereas several simulations of homogeneous nucleation using the mW coarse-grained water model\cite{Molinero2009} have been reasonably successful,\cite{Li2011, Reinhardt2012, Moore2010, Moore2011b} all-atom simulations have been less so. The mW potential is a good representation of the structure and the thermodynamics of water; however, it has unrealistically fast dynamics and no representation of hydrogens, and so the use of an all-atom model would offer considerable further insight into the process. However, Matsumoto and co-workers' single MD trajectory of TIP4P water nucleating into ice remains the only successful brute-force simulation of homogeneous ice nucleation with an all-atom model.\cite{Matsumoto2002} Other simulations have used small system sizes or looked at conditions that are not representative of homogeneous nucleation from the bulk liquid water. While brute-force simulations of a rare event are unlikely to be successful, the use of rare event methods can potentially allow us to compute free energy landscapes for nucleation. Such calculations have recently been attempted in the homogeneous ice nucleation simulations of Radhakrishnan and Trout, who used umbrella sampling,\cite{Radhakrishnan2003b,Radhakrishnan2003} and Quigley and Rodger, who used metadynamics.\cite{Quigley2008} However, the use of global order parameters in driving homogeneous ice nucleation can lead to non-physical nucleation pathways, as we discuss below.

It is a surprising state of affairs that modern simulation methods have so far not been able to capture the fundamental physical behaviour of the homogeneous nucleation process of ice. Some of the outstanding problems are how to simulate the homogeneous nucleation of ice using a local measure of order, and the determination of a free energy landscape using such an order parameter. Here, we address one of these aspects: we develop some appropriate order parameters to allow us to drive the nucleation process and grow a single ice cluster using all-atom models of water.  We first examine order parameters used in nucleation studies in general (Section~\ref{sect-ord-params}) and then introduce the order parameters we use in all-atom water model simulations  (Section~\ref{sect-ord-params-ice}). We discuss the water potential and simulations methods we used in Section~\ref{sect-simulations}, and we present the results of driving nucleation in Section~\ref{sect-results}. Finally, we discuss the outstanding problems that need to be overcome in order to obtain a free energy landscape and nucleation rates in Section~\ref{sect-conclusions}.

\section{Order parameters in nucleation studies}\label{sect-ord-params}
In order to monitor the process of nucleation, we require a quantitative measure that can distinguish how far along the process is. The quantity describing this is usually known as an order parameter. The first step in deciding on an order parameter is to classify particles as being solid-like or liquid-like in nature. In the nucleation literature, such classification is often based on the Steinhardt classification parameter\cite{Steinhardt1983,TenWolde1996}
\begin{equation}
q_{l}(i) = \left[ \frac{4\uppi}{ 2l+1}  \sum_{m=-l}^{+l} \abs{q_{lm}(i)}^2 \right]^{1/2},
\end{equation}
where
\begin{equation}
q_{lm}(i) =  \frac{1}{N_\text{neighs}(i)} \sum_{j=1}^{N_\text{neighs}(i)} Y_{lm}\left(\theta_{ij},\,\varphi_{ij}\right),\label{sph-expansion-coeffs}
\end{equation}
$Y_{lm}\left(\theta_{ij},\,\varphi_{ij}\right)$ are the spherical harmonics, $\theta$ and $\varphi$ are the polar angles measured in an arbitrary laboratory frame of reference and $N_\text{neighs}(i)$ is the number of neighbours of particle $i$.\cite{Note1} It is important to note that all $q_l(i)$ are rotationally invariant regardless of the choice of $l$. There is no radial component in this scheme; one can be introduced if necessary. However, a limited radial dependence arises through our definition of neighbours. In nucleation studies, we are often hoping to compare the environment about a particle to a symmetrical crystalline system, and so we can simply pick the value of $l$ that best corresponds to the symmetry of the crystalline system and compute the spherical harmonic expansion coefficients for only that $l$.\cite{Steinhardt1983} It is both convenient and computationally less expensive to replace the complex spherical harmonics with their real analogues;\cite{Blanco1997, Reinhardt2012} we continue to denote complex conjugates in the following for generality, but they can be dropped if real spherical harmonics are used.  

\subsection{Global order parameters}
In many studies, the local Steinhardt classification parameters (Eq.~\eqref{sph-expansion-coeffs}) are averaged across the system to give global Steinhardt order parameters; these are typically expressed as the magnitude of the vector sum of the local classification parameters averaged over all $N$ particles in the system, namely\cite{Steinhardt1983}
\begin{equation}Q_{l} = \left[\frac{4\uppi}{2l+1} \sum_{m=-l}^{+l} \abs{ \frac{1}{N_\text{tot}} \sum_{i=1}^{N} N_\text{neighs}(i) q_{lm}(i) }^2 \right]^{1/2},\end{equation}
where $N_\text{tot}=\sum_{i=1}^N N_\text{neighs}(i)$. Although local classification parameters $q_l$ are a good measure of the local order about a particle, they are generally non-zero both in the solid phase and in the liquid phase, as the liquid is often reasonably well-ordered, especially when considering only its first neighbour shell. However, the vectors add incoherently in the liquid phase and the global order parameter $Q_l$ averages out to zero for large systems, whilst it does not do so in the solid phase.\cite{TenWolde1996} As a result, the increase in the magnitude of $Q_l$ can be used to track how solid-like a system is. Global order parameters of this type have been used in two previous studies of homogeneous ice nucleation.\cite{Quigley2008,Radhakrishnan2003b} However, global order parameters are not ideal in nucleation studies,\cite{TenWolde1996} particularly in studies where the system is not only tracked, but driven to increase the value of a global order parameter.\cite{Reinhardt2012}

First of all, in nucleation studies, we often wish to perform calculations in such a way as to enable us to compare the results to classical nucleation theory. To do this, we need to know the size of the largest crystalline cluster, usually by knowing how many particles there are in the cluster. However, with a global order parameter, we have no knowledge of what the size of the largest crystalline cluster is; indeed, the physical meaning of any particular value of a global order parameter is not only system-size dependent, but physically hard to interpret, and a free energy landscape calculated as a function of a global order parameter does not have a clear physical interpretation in terms of nucleation.

Secondly, the interfacial free energy is in competition with the more favourable entropy arising from a larger number of smaller clusters. It can be shown\cite{TenWolde1996} that this entropy can play a significant r\^{o}le in small systems that can be simulated on computers: in the early stages of nucleation, many small nuclei are always more favourable than one single nucleus comprising the same number of particles. There is, however, a crossover to the expected behaviour once a certain nucleus size has been passed. This means that, when comparing results to classical nucleation theory, a `global' measure of crystallinity -- which effectively induces an entropic break-up of small clusters -- is inappropriate.\cite{TenWolde1996}

Finally, we have suggested in our previous work\cite{Reinhardt2012} that the pathways produced when the nucleation process is driven by global order parameters may be inconsistent with the natural nucleation pathways (\textit{i.e.}~those pathways that would occur for an unbiassed system given sufficient time), particularly so in the case of ice. Since there is no distinction between the liquid and the solid states of particles when global order parameters are used, driving the system to increase its global order parameter can potentially induce an orientational coherence even in the liquid state. This suggests that a particle is influenced not only by its neighbours, but potentially by a crystalline cluster that is very far removed from it, even though such a long-range interaction has no basis in reality.\cite{Reinhardt2012} Furthermore, when rare event techniques are applied to a system, it is relatively easy to compensate for arbitrarily large free energy barriers; there is a danger, therefore, that if the natural, lowest free energy pathway is dynamically slow (having accounted for the free energy barrier associated with the process itself), such a pathway may not be observed and a higher free energy pathway could be found instead provided that it is dynamically faster and its higher free energy has been negated by a rare event method. We have previously also suggested that, for ice nucleation, the use of global order parameters in driving nucleation may lead to precisely such high free energy pathways.\cite{Reinhardt2012}

\subsection{Local order parameters}
Even though the local Steinhardt classification parameters defined above cannot distinguish between solid and liquid particles on their own, there are a few approaches that allow us to do this without sacrificing their local nature. It is often convenient to calculate the dot products of the individual local classification parameters expressed in vector form, $\mathbold{q}_l(i)$, whose $(2l+1)$ components are the Steinhardt parameters $q_{lm}(i)$ for $m\in[-l,\,l] \cap \mathbb{Z}$, with the equivalent vectors of a particle's neighbours.\cite{TenWolde1996} We calculate the rotationally invariant function $d_l(i,\,j)= \hat{\mathbold{q}}_{l}^{\phantom{\star}}(i) \cdot \hat{\mathbold{q}}_{l}^{\star}(j)$, where $i$ and $j$ are neighbours; this dot product value ranges between $-1$ and $+1$. By plotting the distribution of $d_l$ values for the liquid and the crystalline phases, a critical threshold $d_\text{c}$ can be determined as the first point where the probability of being in the solid phase is non-zero.\cite{
TenWolde1996} The number of crystalline connections is then defined as\cite{TenWolde1996}
\begin{equation}n_\text{connections}(i) = \sum_{j=1}^{N_\text{neighs}(i)}  \operatorname{H}(d_l(i,\,j) - d_\text{c}), \end{equation}
where $\operatorname{H}$ is the Heaviside step function. The number of connections should be higher in the solid phase than in the liquid phase, and a criterion involving a threshold minimum number of connections to distinguish between the two phases is often a good classification parameter.\cite{TenWolde1996, Wedekind2007}

Another procedure involves calculating the neighbour-averaged contribution,\cite{Lechner2008,Jungblut2011}
\begin{equation}\avg{q_{lm}(i)} = \frac{1}{N_\text{neighs}(i)+1} \sum_{j=0}^{N_\text{neighs}(i)}  q_{lm}(j),\end{equation}
where $j$ runs over all the neighbours of particle $i$, and includes the particle itself (when $j=0$). The average local bond classification parameter is then given by
\begin{equation}\avg{q_l(i)} = \left[\frac{4\uppi}{2l+1} \sum_{m=-l}^{+l} \abs{\avg{q_{lm}(i)}}^2\right]^{1/2}.\end{equation}
In both approaches mentioned, the second neighbour shell is effectively taken into account through the use of local Steinhardt vectors of the first neighbour shell, either by averaging or by taking dot products.

Any two particles belong to the same crystalline cluster if they are both classified as being solid-like and are located within a certain fixed distance of one another (that is, they are neighbours). Once all the particles have been classified, the size of the largest such cluster is normally calculated in nucleation studies; this then acts as the overall (local) order parameter.

It is very useful for an order parameter used in driving nucleation to be local in nature; however, that an order parameter is local is not sufficient for it to be a valid metric used for driving the nucleation process. For example, in their ice nucleation simulations,\cite{Brukhno2008, Malkin2012} Brukhno and co-workers used a maximum director projection approach to yield local order parameters. However, although these order parameters do permit the growth of ice to be driven in a fixed orientation with respect to the simulation box, the rotational bias inherent in the procedure induces a non-local and non-physical orientational coherence in the growing ice cluster; we suggest that, as a result, this order parameter may not be suitable to study ice nucleation.

More complex order parameters used to track larger molecule crystallisation have also been proposed,\cite{Santiso2011} whilst an overview of many simpler order parameters used in vapour-liquid nucleation was produced by Senger and co-workers.\cite{Senger1999}

\section{Order parameters for homogeneous ice nucleation}\label{sect-ord-params-ice}
\begin{figure}[tb]
\begin{center}
\includegraphics{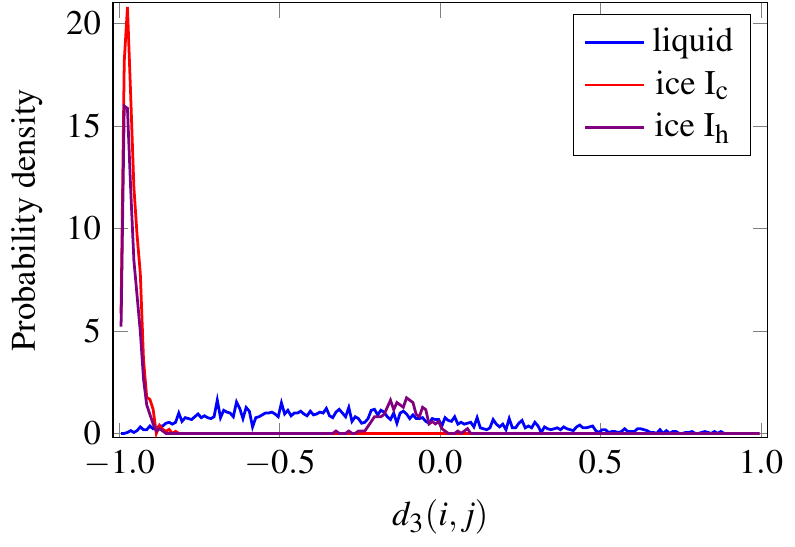}
\end{center}
\caption{A typical probability density distribution for all pairs of \protect{$d_3(i,\,j) = \hat{\mathbold{q}}_3^{\protect\phantom\star}(i)\cdot\hat{\mathbold{q}}_3^{\star}(j)$}, where the centres of mass of molecules $i$ and $j$ are within 3.5\;\AA\ of each other. The three states depicted were equilibrated at 200\;K (using the TIP4P/2005 water model) and the ice structures are not, therefore, `perfect'. This figure is analogous to those in Refs~\onlinecite{Moore2010b,Romano2011,Reinhardt2012}.}\label{fig-histogram-dotProds-3-all}
\end{figure}
The choice of an order parameter to drive ice nucleation is not trivial: as discussed above, it is preferable that it be local; it must be forgiving enough to be able to induce the growth of a small ice cluster; and it must be strict enough to ensure that the structure grown is actually ice-like and has, ultimately, long-range order. In our simulations, we use a variation of the dot product approach described above that is commonly used in studies of tetrahedral liquids.\cite{Ghiringhelli2007,Moore2010b,Romano2011,Reinhardt2012} We choose to use $l=3$, since the $l=3$ spherical harmonics are the ones best describing tetrahedrality. A plot of the distribution of $d_3(i,\,j)$ is shown in Fig.~\ref{fig-histogram-dotProds-3-all}; to account for the eclipsed bond in hexagonal ice, we define a classification parameter as
\begin{equation}n_\text{connections}(i) = \sum_{j=1}^{N_\text{neighs}(i)}  \Gamma(d_3(i,\,j)),\end{equation}
where
\begin{equation}\Gamma(x) = \begin{cases} 1  \qquad & \text{if } [(x < -0.825) \lor (-0.23 < x < 0.01)] , \\ 0 & \text{otherwise}. \end{cases}
\end{equation}
These limiting values were chosen to encompass $d_3(i,\,j)$ regions (Fig.~\ref{fig-histogram-dotProds-3-all}) where the probability density function for either ice phase has a value greater than 0.1. We classify a molecule as ice-like if $n_\text{connections} \ge 3$ and as liquid-like otherwise. This gives perfect identification in both equilibrated cubic and equilibrated hexagonal ice. The order parameter we use to track the progress of nucleation is the size of the largest cluster of molecules classified as ice, where two molecules belong to the same crystalline cluster if they are both ice-like and their centres of mass are within 3.5\;\AA\ of each other.

\begin{figure}[tb]
\begin{center}
\includegraphics{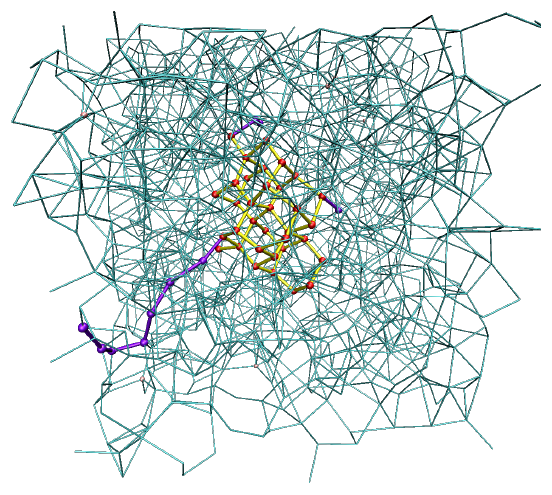}
\end{center}
\caption{An example of non-ice-like chain growth in TIP4P/2005 umbrella sampling simulations when using the order parameter without chain removal as described in the text. The system has 1000 molecules at 240\;K, starting from a 24-molecule cluster. Molecules classified as being part of the largest crystalline cluster are shown in red and violet; there are 45 molecules in this cluster. Molecules whose centres of mass are within 3.5\;\AA\ are connected with lines. Molecules shown in violet would be removed from the largest crystalline cluster on application of the chain removal algorithm described in the text.}\label{fig-TIP4Pnucl-origOrdParam-chainGrowth}
\end{figure}

Unfortunately, whilst this classification procedure (regardless of the precise details of the parameterisation of the limiting values) works well for the mW potential,\cite{Reinhardt2012,Moore2010b} it does not do so in all-atom models of water. When used in the form presented above in umbrella sampling\cite{Torrie1977} or forward flux sampling simulations,\cite{Allen2006b} natural fluctuations in the system often result in molecules satisfying the order parameter even if they are not really ice-like. When forced to grow with a biassing potential, `chains' form more easily than real ice grows, even though such chains actually represent an abuse of the order parameter, and `ice' structures as depicted in Fig.~\ref{fig-TIP4Pnucl-origOrdParam-chainGrowth} are commonplace. The real issue is not just that such chains form, but that when they do form, the system is not subsequently able to transform to the correct (compact) ice structure, and ice growth is arrested. For example, in forward flux sampling simulations, the probability of reaching the next interface along the reaction co-ordinate rapidly approaches zero once the system exhibits predominantly chains, and in umbrella sampling simulations, the system is frustrated so much that it fails to grow further even when using extremely large biassing potentials. This suggests that chain growth of this type is not a natural feature of ice nucleation.

Such chain growth can be observed with all variants of local order parameters we have tried. In order to alleviate the problem of chain growth, we (a) classify any molecule with more than four neighbours (within 3.5\;\AA) as being liquid, and (b) explicitly exclude molecules belonging to chains from the largest crystalline cluster. We achieve the latter by removing any molecules with only one neighbour belonging to the largest cluster from the largest cluster, except if that single neighbour is connected to three further molecules in the largest cluster. This allows `chains' comprising a single molecule to form and thus allows ice to grow. We iterate the procedure until no further molecule is removed.

Removing chains could be problematic in the initial stages of nucleation: it is impossible for an ice structure smaller than a single chair (or boat) not to be formed of chains, and small rings may form instead if forced. However, due to the similarity of liquid water and ice, it is possible to wait for a boat or chair cluster to form spontaneously in a simulation, and limit umbrella sampling to systems with clusters larger than $\sim$10 molecules. Doing so does not appear to be a significant limitation when the critical cluster is expected to comprise over 100 molecules.

\begin{figure}[tb]
\begin{center}
\includegraphics{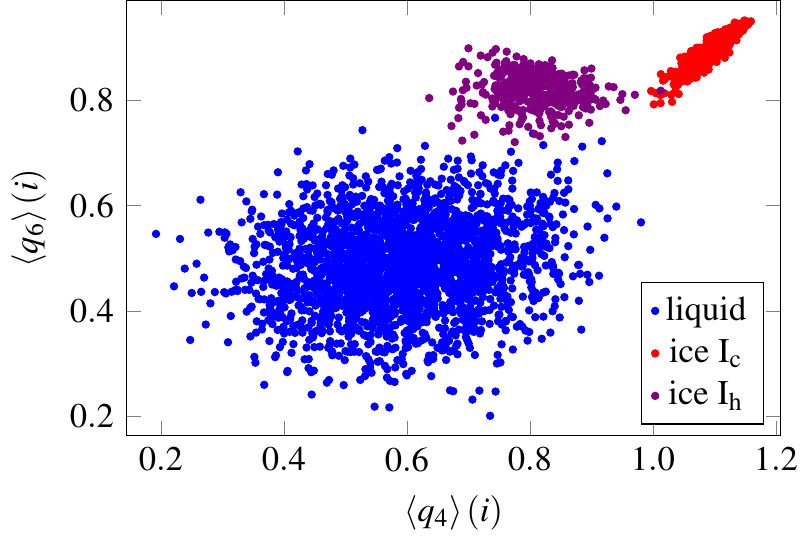}
\end{center}
\caption{Neighbour-averaged order parameters for systems of ices I$_\text{h}$ and I$_\text{c}$ and liquid water. All systems were equilibrated at 200\;K using the TIP4P/2005 water model, and they contain different numbers of molecules. The neighbour cutoff distance was 3.5\;\AA.}\label{fig-scatter-Lechner-TIP4P2005}
\end{figure}

Although neighbour-averaged classification parameters are excellent at distinguishing between the phases of systems without much structure in the liquid phase,\cite{Lechner2008,Lechner2011} it is less clear whether the same applies to well-structured liquids like water. Using $l=4$ and $l=6$ (as depicted in Fig.~\ref{fig-scatter-Lechner-TIP4P2005}) gives better separation between the phases than does using $l=3$. Liquid water and hexagonal ice are less well-separated in the $\avg{q_4}$-$\avg{q_6}$ plane than in the Lennard--Jones case; nonetheless, a choice of $\avg{q_6(i)}>0.7$ as an ice-liquid boundary would appear to be reasonable. Calculating the size of the largest cluster with this method results on average in only slightly smaller clusters compared to those resulting from the dot product approach, and the formation of chains is as problematic as when using the latter. We have shown\cite{Reinhardt2012} that for the mW monatomic model of water,\cite{Molinero2009} a dot product approach leads to a free energy profile that is almost entirely consistent with classical nucleation theory. By contrast, including `surface' molecules in the largest ice cluster, as attempted in other studies with the same model of water,\cite{Li2011, Moore2011} appears to reduce the agreement with classical nucleation theory.\cite{Reinhardt2012} Lechner and co-workers have recently used a combination of dot product vectors and their neighbour-averaged classification parameters to study the r\^{o}le of the surface and the bulk terms when comparing simulation data to classical nucleation theory for a soft-core colloid model.\cite{Lechner2011, Lechner2011b} Using neighbour-averaged classification parameters reduces the size of the critical cluster when compared to one calculated using a dot product approach in their work.\cite{Lechner2011,Lechner2011b} The clusters we have analysed with both approaches show only a small difference in cluster size; on average, the neighbour-averaged clusters are slightly smaller, but for individual configurations, the converse can also hold. Nonetheless, given that the number of molecules classified as being part of the cluster by the dot product approach is itself a rather conservative estimate of the cluster size,\cite{Reinhardt2012} and since it would be computationally extremely expensive to perform two-dimensional umbrella sampling with an additional order parameter, we restrain ourselves, for the time being, to using the dot product approach only. However, the neighbour-averaged approach is an attractive alternative, and ensuring that clusters are of essentially the same size with both approaches is a useful confirmation that the exact details of the order parameter do not appear to change the outcome significantly.

\section{Simulation details}\label{sect-simulations}
An empirical model that reproduces a large number of experimental results at a reasonable computational cost is the TIP4P/2005 model,\cite{Abascal2005} which seems to be the best of the `simple' all-atom models available at present, and is one that works well across several phases\cite{Vega2009,Vega2011,Kiss2011} and at large supercoolings.\cite{Pi2009} Although more complex models can, at a considerable computational expense, capture more of the underlying physical behaviour, their use, at least in their present state of development, does not necessarily result in a better description of water.\cite{Huggins2012, Gladich2012, Habershon2011b, Tironi1996,Gonzalez2011} In this work, we therefore use the TIP4P/2005 model of water, as described in the original paper by Abascal and Vega,\cite{Abascal2005} using the same parameters for cutoffs and Ewald summation.

To simulate the nucleation of ice, we use the Metropolis Monte Carlo (MC) approach\cite{Metropolis1953} in the isobaric-isothermal ensemble, coupled with umbrella sampling\cite{Torrie1977} to drive the process. In umbrella sampling, an additional term dependent on the order parameter arises in the Boltzmann factor when considering whether to accept or reject a trial move; whilst this additional energy is often implemented as a quadratic bias potential,\cite{Auer2004} in our implementation, we use adaptive umbrella weights.\cite{Mezei1987} We typically choose these weights to correspond roughly to the negative of the free energy predicted by classical nucleation theory for a given cluster size: the precise potentials we used were changed slightly as the simulations progressed.\cite{Mezei1987}

All-atom models of water have particularly slow dynamics, especially for crystal growth, making any possible improvement in computational speed worth considering. It has been shown that ice crystal growth occurs more rapidly (in computer time) in molecular dynamics (MD) simulations than in corresponding Monte Carlo ones,\cite{Fernandez2006, Conde2008} which may suggest that the collective motion possible in MD simulations helps to speed up the dynamics of cluster reorganisation, and thus aids us in driving crystallisation. The choice of MD simulations over MC simulations is therefore appealing; however, as we are ultimately interested in the free energy landscape of ice nucleation, we wish to continue to use umbrella sampling and thus Monte Carlo simulations.

While the majority of simulations presented in this paper used the standard Monte Carlo method, we have recently begun to couple Monte Carlo with MD simulations in a hybrid Monte Carlo approach,\cite{Duane1987, *Heermann1990, *Mehlig1992,*Brass1993,*Tuckerman2010} where short MD simulations replace single particle rotational and translational Monte Carlo moves. Provided that the MD integrator is time reversible and symplectic and that the choice of momenta from the Maxwell--Boltzmann distribution is accounted for in the Metropolis acceptance criterion, detailed balance is obeyed\cite{Duane1987, *Heermann1990, *Mehlig1992,*Brass1993,*Tuckerman2010} irrespective of the fact that the MD hamiltonian does not incorporate an umbrella sampling term. We implement the symplectic and time reversible quaternion-based algorithm of Miller III and co-workers\cite{Miller2002, Kamberaj2005} to simulate rigid body rotations.\cite{Note2,Miyamoto1992}  We are able to drive nucleation using both methods; however, simulations are considerably faster in real time when the hybrid Monte Carlo approach is used,\cite{Note3} which confirms the importance of collective motion for nucleation.

\section{Nucleation pathways}\label{sect-results}
We have run umbrella sampling simulations in a variety of systems with TIP4P/2005 water. Typical simulations involved between 1900 and 2500 water molecules and three distinct scenarios were considered: growth from a seed hexagonal ice cluster, growth from a seed cubic ice cluster and growth directly from the supercooled liquid water.

\begin{figure*}[tbp]
\begin{center}
\includegraphics{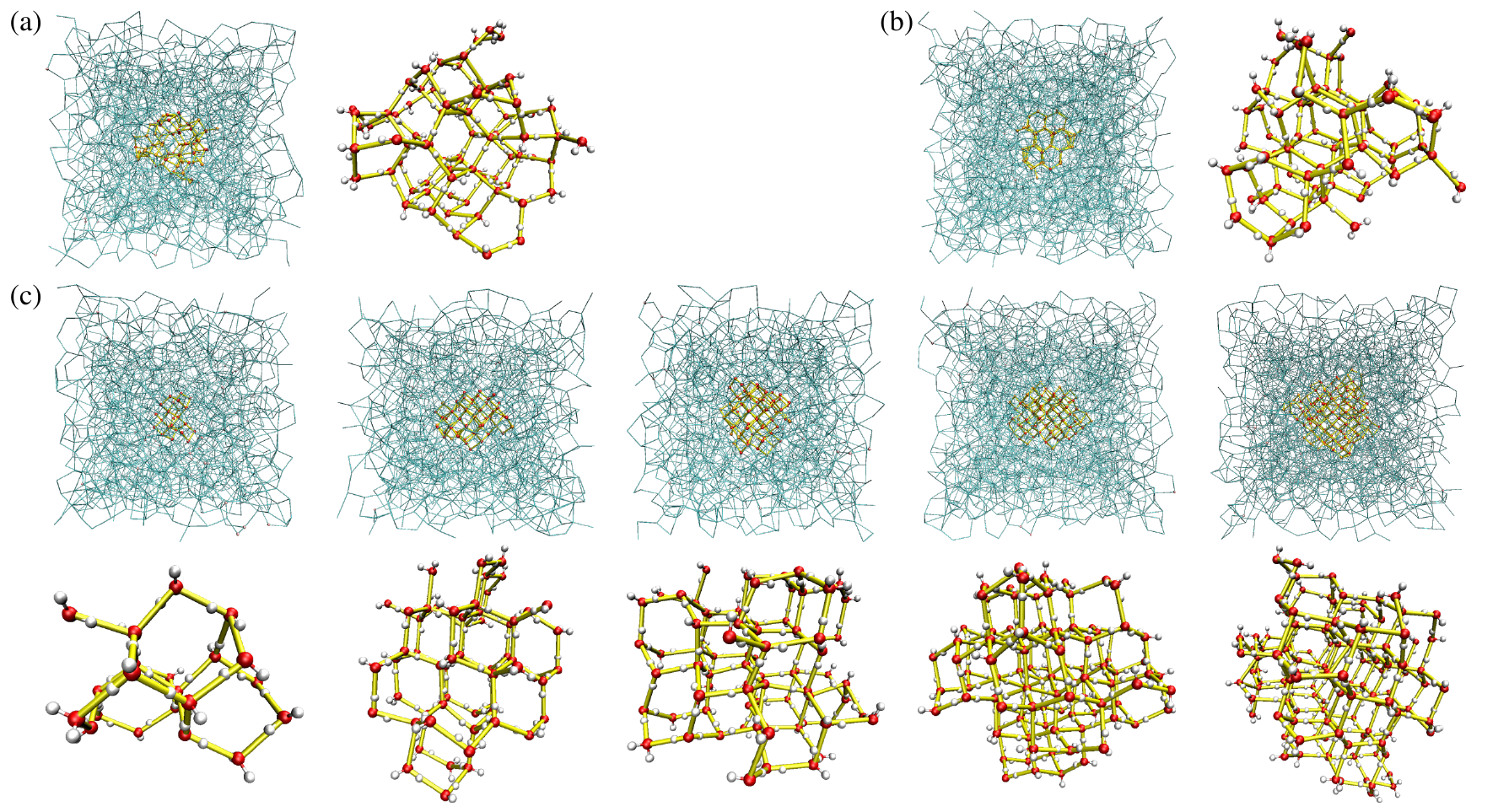}
\end{center}
\caption{Representative nucleation snapshots from umbrella sampling simulations of TIP4P/2005 water. In each case, two pictures depict the same cluster from different perspectives; one within the liquid framework (in cyan) and one showing solely the largest crystalline cluster. In the former, spheres represent centres of mass of molecules classified as ice: red spheres correspond to cubic ice, orange spheres correspond to hexagonal ice and pink spheres correspond to ice molecules not within the largest crystalline cluster. Pictures representing solely the largest cluster depict both the oxygen (red) and the hydrogens (white) of each molecule. In (a), an 82-molecule ice cluster grown from the supercooled liquid at 240\;K is shown; in (b), a 73-molecule ice cluster grown from a small cluster of I$_\text{h}$ ice  at 240\;K is shown; and in (c), a series of ice clusters of increasing size (comprising 23, 60, 77, 107 and 145 molecules from left to right) grown from a small cluster of I$_\text{c}$ ice  at 200\;K is depicted. There are 1900 molecules in the system in (a) and the first three configurations of (c), and 2500 molecules in (b) and the last two configurations of (c). Simulations of nucleation from a hexagonal seed (shown in (b)) were undertaken using the hybrid Monte Carlo approach, and the rest by a standard Metropolis Monte Carlo approach. $p=\text{1\;bar}$.\label{fig-TIP4P-forced-nucl-snapshots}}
\end{figure*}

The approach we used was to bias the umbrella sampling weights to favour larger clusters, although the weights chosen were only slightly overbiassed compared to the classical nucleation theory prediction of the free energy barrier. If the umbrella weights bias the growth to be too quick, then the system can begin to grow defective crystal nuclei that cannot repair themselves by shrinking and regrowing: the weights must be sufficiently small to allow clusters to grow and shrink throughout each umbrella sampling window.

Several snapshots of ice growth in such simulations are depicted in Fig.~\ref{fig-TIP4P-forced-nucl-snapshots}. Provided that umbrella sampling does not attempt to drive the nucleation too quickly, the resulting ice clusters appear to be reasonable: for example, they do not span the simulation box and are compact. This suggests that the order parameter presented above is a suitable order parameter both to track and to drive the process of homogeneous ice nucleation.

It is intriguing to note that it appears that in what we believe to be overbiassed driving of nucleation,  cubic ice seeds seem to grow in a cubic fashion, and conversely, hexagonal ice seeds result in the growth of hexagonal ice. Growth directly from the supercooled liquid is somewhat more tricky: because the order parameter we use cannot track clusters smaller than 6 molecules, and cannot readily track the growth of very small clusters of ice, it was necessary to wait for small clusters to form spontaneously. These were then biassed to grow further, although with a very gentle set of umbrella weights, and only when the growth became spontaneous with that set of weights did we progress to higher umbrella sampling windows. Analogously to what we observed in mW simulations,\cite{Reinhardt2012} ice grown directly from the supercooled liquid contains both cubic and hexagonal ice patterns, with cubic ones dominating, but less so than in the corresponding mW clusters: for example, 60-molecule ice clusters in this work were classified to have approximately 70\;\% core cubic ice, whilst the mW analogues were about 90\;\% cubic. Whether this is a result of a true difference between the TIP4P/2005 and mW models of water or simply a consequence of overbiassed non-equilibrated driving in this work is unclear and warrants further investigation. The ice clusters observed in this work are roughly spherical, and this sphericity follows the same trends as for the mW model nucleation reported previously.\cite{Reinhardt2012}

Matsumoto and co-workers looked at some properties of the clusters they observed along the nucleation pathway of their spontaneous MD nucleation trajectory.\cite{Matsumoto2002} Their clusters comprise molecules that are connected by a network of long-lived hydrogen bonds, and thus include the majority of the cluster surface. Nevertheless, non-compact clusters were observed not to lead to successful nucleation pathways, whereas the clusters associated with nucleation were much more compact and exhibited only few chains (Fig.~4 of Ref.~\onlinecite{Matsumoto2002}). We can compare this behaviour to what we observe in driven nucleation simulations. Although the order parameter we use to drive nucleation does not count molecules in chains as belonging to the largest crystalline cluster, it does not, in principle, suppress such growth, and so if chain growth were a natural feature associated with nucleation, we might expect chains to grow nonetheless. In order to investigate whether chains are such a feature of the nucleation process or, as we suggested previously, an artefact of the order parameter coupled with slow dynamics of ice growth, we have calculated the numbers of molecules that are classified as belonging to chains along the nucleation pathway as driven by the order parameter presented above. In fact, chains are usually present on the surface of the growing ice nuclei; however, the absolute numbers are small: for example, in the set of simulations started from a hexagonal seed cluster, a cluster of 50 molecules has on average only 3.25 molecules belonging to chains, and such chains grow reasonably uniformly on all sides of the crystalline cluster surface. Although certain configurations do exist with longer chains, chain growth does not seem to play a crucial r\^{o}le in ice nucleation as driven by our order parameter, which is consistent with the behaviour observed in spontaneous nucleation by Matsumoto and co-workers.

\begin{figure}[tb]
\begin{center}
\includegraphics{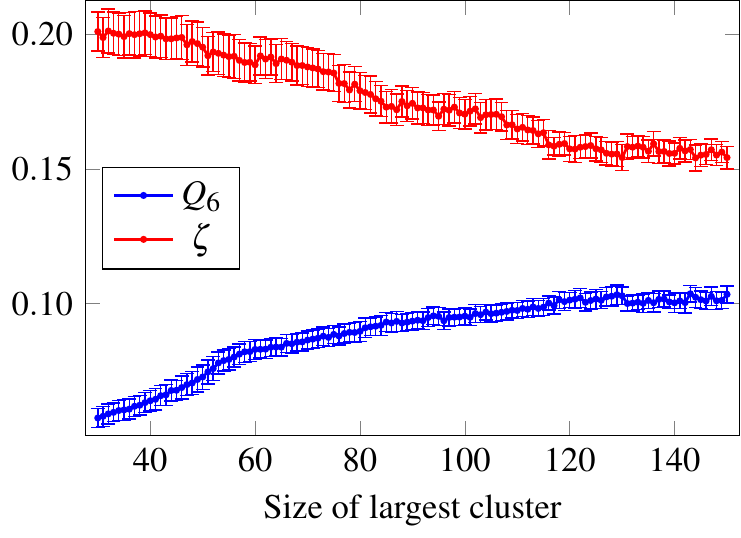}
\end{center}
\caption{The global order parameters $Q_6$ and $\zeta$ calculated as a function of the size of the largest crystalline cluster, the order parameter used to drive nucleation in this work, for the system seeded with a cubic ice nucleus. Error bars show the standard deviation for the population of configurations at each cluster size. The results depicted here refer to the 576 particles nearest the centre of mass of the ice nucleus. $T=\text{200\;K}$, $p=\text{1\;bar}$.\label{fig-TIP4P-forced-ordParamsOfGrowth}}
\end{figure}

Although we made no serious attempt to equilibrate the nucleation simulations, it may nevertheless be possible to gain physical insight into the nucleation pathway from our simulations. In order to compare the pathway observed with that reported in our simulations using the mW potential\cite{Reinhardt2012} and with Quigley and Rodger's simulation of TIP4P nucleation,\cite{Quigley2008} we calculate the Steinhardt-style $Q_6$ and Chau--Hardwick-style\cite{Chau1998,*Errington2001} tetrahedrality parameters as defined by Quigley and Rodger,\cite{Quigley2008} including their smoothing function. These are given by
\begin{equation}
 Q_l = \left( \frac{4\uppi}{2l+1} \sum_{m=-l}^{+l} \left| \frac{1}{4N}  \sum_{i=1}^{N} \sum_{j\ne i}^{N} f(r_{ij}) Y_{lm}\left(\theta_{ij},\,\varphi_{ij}\right)  \right|^2  \right)^{1/2},
\end{equation}
where $N$ is the number of particles and $f(r_{ij})$ is the smoothing function, and
\begin{equation}
 \zeta =  \frac{1}{4N} \sum_{i=1}^N \sum_{\substack{j=1\\j\ne i}}^N \sum_{\substack{k>j\\k\ne i}}^{N} f(r_{ij})f(r_{ik}) \left( \hat{\mathbold{r}}_{ij} \cdot \hat{\mathbold{r}}_{ik} + 1/3 \right)^2,\label{equation-ChauHardwick}
\end{equation}
where $\hat{\mathbold{r}}_{ij}$ is the unit vector from particle $i$ to particle $j$. The smoothing function is defined as
\begin{equation}
f(r) = \begin{cases}
       1 & \quad \text{if } r \le \text{3.1\;\AA}, \\
       \left(\cos\frac{(r/\text{\AA}-3.1)\uppi}{0.4}+1\right)/2 & \quad \text{if } \text{3.1\;\AA} < r \le \text{3.5\;\AA}, \\
       0 & \quad \text{otherwise.}
    \end{cases}
\end{equation}
We take into account the nearest 576 molecules from the centre of mass of the largest crystalline cluster as determined by the local order parameter defined above in order to ensure that these results can be compared to the previous work. The resulting diagram for the simulation of nucleation from a cubic ice seed nucleus is depicted in Fig.~\ref{fig-TIP4P-forced-ordParamsOfGrowth}. Although the curves are rather noisy, as can be expected from a set of simulations that have not been equilibrated, and there are clear minor variations in slope corresponding to different umbrella sampling windows, we can nevertheless observe that the two order parameters plotted change roughly linearly as the cluster size increases. This linearity, which reflects the growth of an ice nucleus into a largely unperturbed liquid, is consistent with the nucleation pathway we reported for the mW model of water,\cite{Reinhardt2012} although the actual values of the global order parameters suggest that the system studied here is less well ordered than its mW analogue. This is perhaps not surprising considering that the mW systems we studied previously were very well equilibrated. Importantly, the pathway is rather different from that observed in $Q_6$-$\zeta$ space by Quigley and Rodger,\cite{Quigley2008} as their free energy landscape involves an initial increase in the $Q_6$ orientational order before the tetrahedrality parameter $\zeta$ begins to change, which implies that the entire system, rather than just a crystalline nucleus, becomes more ordered prior to the nucleation event. Since the free energies associated with the nucleation pathway in Quigley and Rodger's study are considerably higher than the ones implied by our (not completely equilibrated) results, this adds further weight to our contention that global order parameters may locate pathways that are not fully consistent with the natural nucleation pathways. It must be emphasised that local order parameters do not necessarily result in more natural reaction co-ordinates; however, if a pathway can be found that has a lower free energy barrier associated with it, then such a pathway will be favoured over one that has a significantly higher free energy barrier.

\section{Discussion and conclusions}\label{sect-conclusions}
We have introduced an order parameter that is capable of tracking and driving the homogeneous nucleation of ice with the TIP4P/2005 water model. The order parameter is local in nature and thus does not exhibit the anomalous behaviour associated with global order parameters. We believe it to be the first rotationally invariant order parameter of this kind that is capable of driving homogeneous ice nucleation in simulations of an all-atom model of water. One of the major difficulties in the development of such an order parameter is that the time that is required to confirm whether an order parameter is fit for purpose is very significant, given that the dynamics at reasonable supercoolings are so slow. In particular, it is important that the umbrella sampling weights not be increased too quickly even if it appears that no growth is forthcoming: one must exercise a considerable degree of patience when performing such nucleation simulations.

\begin{figure}
\begin{center}
\includegraphics{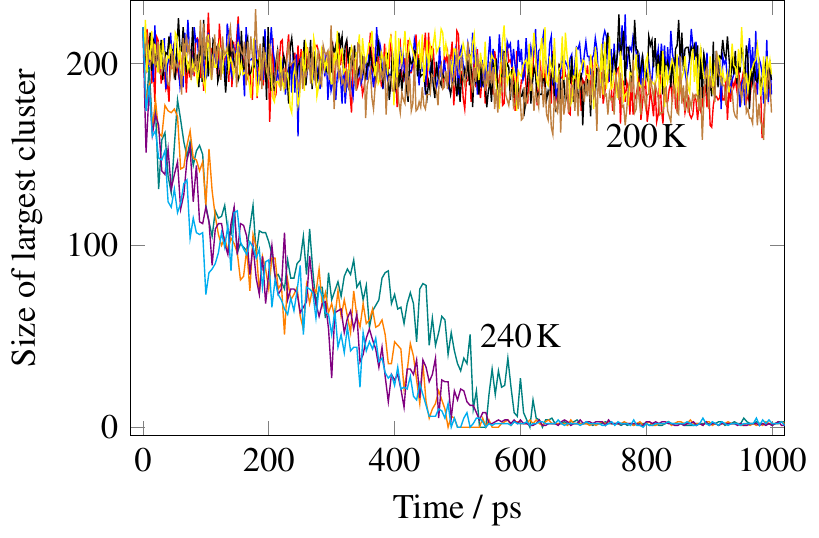}
\end{center}
\caption{MD simulations of melting. The starting point is a crystalline cluster comprising approximately 220 molecules embedded in supercooled liquid water. The curves exhibiting melting were simulated at 240\;K, whilst the remaining ones were simulated at 200\;K. These simulations entailed 2500 TIP4P/2005 water molecules. Note that the melting point of TIP4P/2005 ice is 252\;K.\cite{Vega2011} $p=\text{1\;bar}$.}\label{fig-MD-melting-Q-vs-time}
\end{figure}

The order parameters we have introduced represent the first step in obtaining free energy landscapes and nucleation rates for the homogeneous nucleation of ice from simulations. However, there remain considerable challenges ahead. The basic problem is that the dynamics of ice nucleation for all-atom models are excruciatingly slow,\cite{Fernandez2006, Pan2011} making equilibration very difficult.  This problem is illustrated by Fig.~\ref{fig-MD-melting-Q-vs-time}, which shows that at a reasonable 20\;\% supercooling (200\;K) in the TIP4P/2005 model, a crystalline cluster of approximately 220 molecules in a system of 2500 molecules neither shrinks nor grows. At this temperature, classical nucleation theory would predict that a 220-molecule crystalline cluster is post-critical and so we might expect to see it grow. However, using a cluster that classical nucleation theory would predict to be pre-critical in size does not alter the system's static behaviour.  Although only the first 1\;ns of this pure MD simulation is shown in Fig.~\ref{fig-MD-melting-Q-vs-time}, the results are no different when simulated up to 70\;ns, taking approximately a month of CPU time for each trajectory. Furthermore, to allow us to equilibrate a system, we would require not one, but a large number of freezing and melting events to be able to be simulated in such a period of computer time.

One control parameter at our disposal is, of course, the temperature. The fastest rate of growth of an ice-water interface is generally some 10\;K below the freezing point,\cite{Pereyra2011, Weiss2011, Rozmanov2011, Rozmanov2012b} where the increased freezing driving force of cooler systems optimally balances their slower dynamics. The faster dynamics are illustrated by the reasonably rapid melting of the crystalline cluster at 240\;K ($\sim$5\;\% supercooling) as depicted in Fig.~\ref{fig-MD-melting-Q-vs-time}. One consideration that must be borne in mind when choosing a suitable temperature at which to perform simulations is that the majority of experimental rates have been reported at temperatures corresponding to supercoolings of between 27\;\% and 10\;\%;\cite{McDonald1953, *Mossop1955, *Mason1958, *Thomas1952, *Wood1970, *Butorin1972, *Michelmore1982, *Taborek1985, *Stoyanova1994, *Bartell1994, *Huang1995, *Wood2002, *Kabath2006, *Murray2006, *Edd2009, *Stan2009, *Murray2010, *Earle2010,  *Rzesanke2012, *Manka2012} these also encompass the atmospherically relevant conditions. While higher temperature simulations may be easier to run in some ways due to the expedited dynamics, raising the temperature is not without its problems. For example, while we can negate the higher nucleation free energy barrier with umbrella sampling, it is not just the barrier height, but also the critical cluster size  that increases with increasing temperature. Increasing the temperature would thus require us to simulate considerably larger systems than are computationally affordable in order to avoid spurious finite size effects: for illustration, at 200\;K, the critical cluster for TIP4P/2005 water is predicted by classical nucleation theory to encompass $10^2$ molecules, while at 240\;K, this rises to $10^4$ molecules.\cite{Note4,Davidchack2012} We could attempt to extrapolate the free energy barrier to lower temperatures using histogram reweighting\cite{Ferrenberg1988} based on the results of small cluster simulations at higher temperatures. However, the calculation of nucleation rates requires the simulation of critical clusters, and so must be performed at sufficiently low temperatures so that the critical cluster is small enough to be feasible to simulate.

How does one, then, successfully simulate the homogeneous nucleation process and obtain a free energy landscape and nucleation rate? The simplest strategy is simply to wait for a very long time: however, given how computationally challenging the process is, this may involve an inordinate amount of computer time. A second approach is to use more efficient simulation algorithms; indeed, as discussed above, the use of the hybrid Monte Carlo approach gives a significant advantage over standard Monte Carlo simulations in terms of simulation speed. Other tricks of the trade that might be advantageous include the use of hamiltonian exchange\cite{Fukunishi2002} to couple the system of interest to one that is dynamically faster, or the use of reaction fields in place of the computationally expensive Ewald summation.\cite{Huenenberger1998} Finally, the water potential we use is another parameter of the system that is under our control. It has been suggested that there are few differences in the dynamics of ice melting of most common all-atom water models,\cite{Wei2010} and so a possible solution to the equilibration problem may be to use a model of water that is not necessarily the best at describing most experimental properties, but one whose dynamics are computationally faster; examples might include TIP5P(-E)\cite{Mahoney2000, Rick2005} and the Nada--Van der Eerden potential.\cite{Nada2003, Nada2005} Even though such potentials may be computationally more demanding than TIP4P-analogues on a per-step basis, the dynamics of ice growth might nevertheless be faster.\cite{Carignano2007}

In conclusion, the development of a seemingly rigorous order parameter may help us to advance our understanding of ice nucleation; in particular, we have further corroborated our hypothesis that the difficulty in simulating ice nucleation in all-atom models such as TIP4P/2005 is more a result of the slow dynamics of the process rather than of an overwhelmingly large free energy barrier. We hope that the order parameter we have presented here represents a stepping stone towards the successful determination of a free energy landscape and nucleation rate for homogeneous ice nucleation for all-atom water models.

\begin{acknowledgments}
We should like to thank the Engineering and Physical Sciences Research Council and the Direcci\'{o}n General de Investigaci\'{o}n Cient\'{i}fica y T\'{e}cnica (grants FIS2010-16159 and FIS2010-15502) for financial support.
\end{acknowledgments}

\section*{References} 

\end{document}